\documentclass[preprint,twocolumn,10pt,a4paper,byrevtex,prb,showpacs,superscriptaddress]{revtex4}
\usepackage{natbib}
\usepackage{graphicx}
\usepackage{makecell}
\usepackage{textcomp} 
\usepackage{amsfonts}
\usepackage{amsmath}
\usepackage{amssymb}
\usepackage{color}

\begin{document}

\title{Photodoping-driven crossover in low-frequency noise in MoS$_2$ transistors}

\author{Isidoro Martinez}
\thanks{Equally contributing authors}
\affiliation{Depto. Fisica Materia Condensada, C03, Instituto Nicolas Cabrera (INC), Condensed Matter Physics Institute (IFIMAC), Universidad Autonoma de Madrid,28049, Madrid, Spain}
\author{M\'ario Ribeiro}
\thanks{Equally contributing authors}
\affiliation{CIC NanoGUNE, Donostia-San Sebastian 20018, Basque Country (Spain)}
\author{Pablo Andres}
\affiliation{Depto. Fisica Materia Condensada, C03, Instituto Nicolas Cabrera (INC), Condensed Matter Physics Institute (IFIMAC), Universidad Autonoma de Madrid,28049, Madrid, Spain}
\author{Luis E. Hueso}
\affiliation{CIC NanoGUNE, Donostia-San Sebastian 20018, Basque Country (Spain)}
\affiliation{IKERBASQUE, Basque Foundation for Science, Bilbao 48013, Basque Country (Spain)}
\author{F\`elix Casanova}
\thanks{f.casanova@nanogune.eu}
\affiliation{CIC NanoGUNE, Donostia-San Sebastian 20018, Basque Country (Spain)}
\affiliation{IKERBASQUE, Basque Foundation for Science, Bilbao 48013, Basque Country (Spain)}
\author{Farkhad G. Aliev}
\thanks{farkhad.aliev@uam.es}
\affiliation{Depto. Fisica Materia Condensada, C03, Instituto Nicolas Cabrera (INC), Condensed Matter Physics Institute (IFIMAC), Universidad Autonoma de Madrid,28049, Madrid, Spain}

\date{\today }

\begin{abstract}
Transition metal dichalcogenide field-effect transistors (FETs) have been actively explored for low-power electronics, light detection, and sensing. Albeit promising, their performance is strongly limited by low-frequency noise (LFN). Here, we report on the study of LFN in MoS$_2$ FETs on SiO$_2$ substrates in ambient conditions using photodoping. Using this external excitation source allows us to access different non-equilibrium steady states and cross over different noise regimes. We observe a dependence of the noise power spectrum with the transient decay time window, approaching $1/f$-type when the system is closer to equilibrium, and identify a dependence of the LFN with channel thickness. Monolayer/bilayer devices exhibit random telegraph noise for  insulating regimes and $1/f$-type Hooge mobility fluctuations (HMF) for conductive regimes. Thicker devices exhibit mainly $1/f$-type carrier number fluctuations (CNF). In the latter, we observe a photodoping-induced change from a near parabolic to a near linear dependence of the inverse $1/f$ noise amplitude above the threshold gate voltage. This change indicates a crossover in the LFN mechanism from CNF to HMF. We demonstrate that the study of conductance and noise under photodoping is an effective tool to identify dominating carrier noise mechanisms in few-atomic-layer FETs for a wide range of doping regimes.
\end{abstract}

\maketitle

\section{Introduction}
Collective wavelike fluctuations are ubiquitous and inherent to two-dimensional (2D) van der Waals (vdW) semiconductors, being responsible for nontrivial modifications of noncovalent interactions at the nanoscale \cite{Ambrosetti2016}. In few-layer-thick 2D vdW electrical devices, the high surface-to-volume ratio and ultimate thinness of the channel make conduction electrons particularly vulnerable to traps, ionized impurities, and changes in the scattering cross-section. These fluctuations can manifest as low-frequency noise (LFN) with power spectrum $S_I(f)$  closely following an inverse dependence with frequency $f$, $1/f^{\beta}$, where $\beta$ is a characteristic exponent. Although LFN dominates the power spectrum at low frequency, it is the main contributor to the phase noise of sensors and high-frequency operating systems due to its upconversion to high frequencies\cite{Balandin2013}. For these reasons, LFN has emerged as a key limiting factor in the performance of 2D vdW-based devices, particularly under low doping regimes \cite{Sangwan2013,Renteria2014,Kwon2014,Cho2015}. It is thus crucial to understand the processes responsible for the fluctuation of the electrical current in such devices for future applications. 

The unique electrical and optical properties of ultrathin films of transition metal dichalcogenides (TMDs) have been intensively explored in few-atomic-layer field-effect transistors (FETs) \cite{Radisavljevic2015,Sarkar2015}. MoS$_2$ FETs in particular have been actively investigated for low-power electronics \cite{Radisavljevic2015,Wang2012,Kim2012}, light detection, photocurrent generation \cite{Furchi2014,Memaran2015,Buscema2015}, and sensing \cite{Perkins2013,Lin2015}.

In MoS$_2$ FETs, the presence of trapping - de-trapping processes at the channel to insulator and vacuum interfaces has been demonstrated to drive bi-exponential current relaxation\cite{Late2012}, time-dependent contributions to the electron transport characteristics\cite{Guo2015}, and to differences in the carrier density. These processes contribute to the slow relaxation of the photoconductivity in MoS$_2$\cite{Wu2015}.
In recent years, noise in mono-to-multilayer TMD FETs has been mainly described by  carrier number fluctuations, (CNF)\cite{Joo2016,Ko2016,Das2015,Lin2015,Sharma2014,Ghatak2014,Xie2014,Renteria2014,Kwon2014,Na2014,Sharma2015,Rumyantsev2015}, with some exceptional cases reporting phenomenological Hooge mobility fluctuations, (HMF)\cite{Sangwan2013,Na2014}, where $1/f$ noise is interpreted in terms of the fluctuation of the free path length of the charge carriers \cite{Hooge1969}. This percolative nature of the electron conduction has been shown to be a dominant noise mechanism in multilayer WSe$_2$ FETs \cite{Cho2015,Paul2016}. Although CNF is seen as the dominant LFN mechanism in MoS$_2$, there is no clear explanation to the microscopic physical mechanisms behind noise in the cases where HMF has been observed. It is therefore worth investigating photodoping as an external stimulus to access different low-frequency noise regimes in a set of TMD-FETs with different thicknesses.

Here we present a study of electron transport and photoconductivity in monolayer to bulk MoS$_2$ FETs to address three important questions: (i) how transient decays affect LFN in back-gated MoS$_2$ FETs, (ii) whether photodoping, by modulating the charge carrier density, can change the LFN mechanisms, and (iii) if LFN mechanisms in MoS$_2$ FETs undergo a significant change with incremental channel layer number. Furthermore, by providing an additional route to identify the physical mechanisms behind LFN in TMD FETs we pave the road to the development of alternative approaches to optimize TMD-based photodetectors and transistors. 

   \begin{figure}[!h]
   \includegraphics[width=\linewidth]{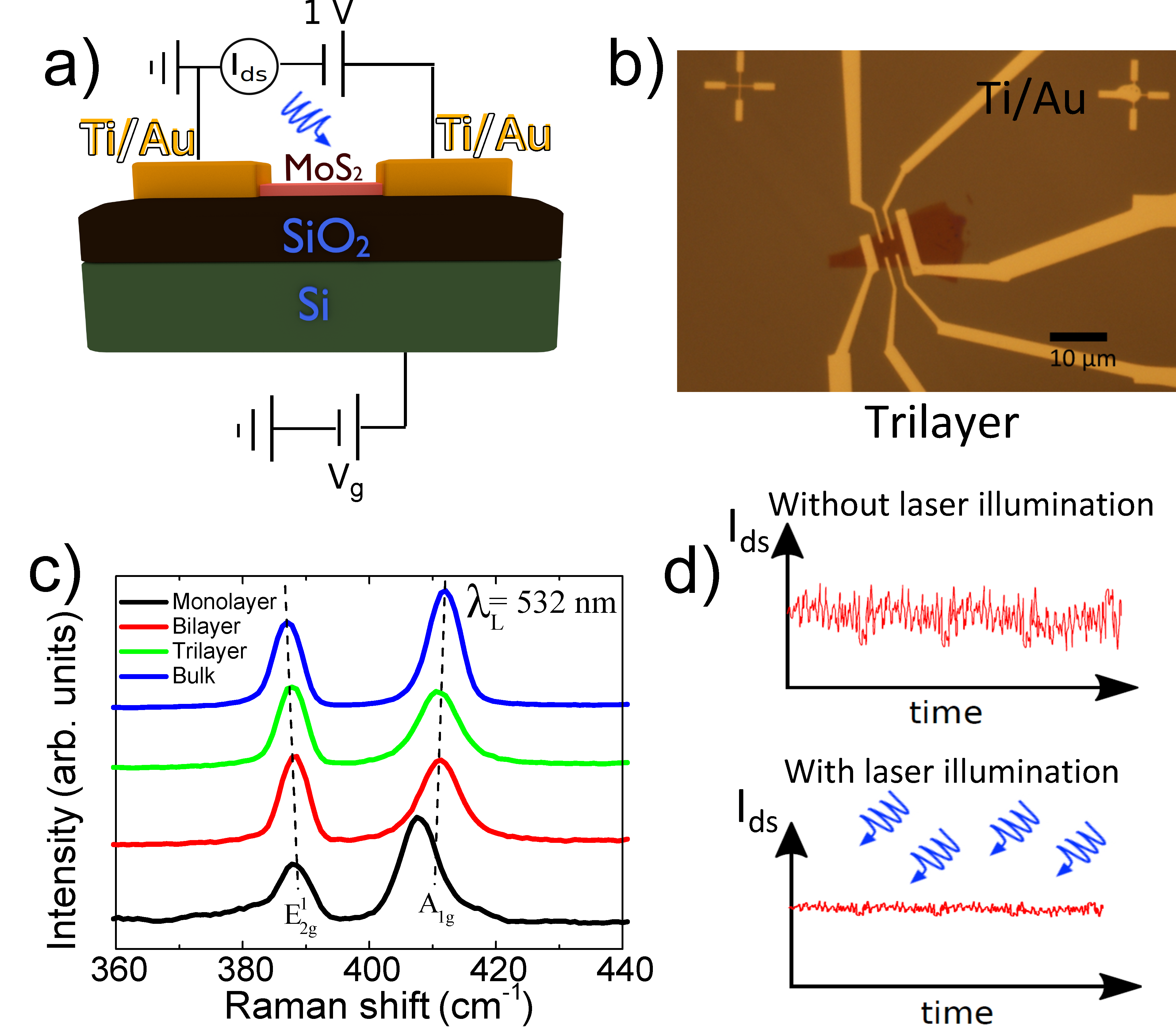}
    \centering
     \caption{a) Sketch of the two-point measurement setup. The MoS$_2$ channel is contacted by Ti/Au contacts, with SiO$_2$ used as backgate dielectric. The open top channel allows for the use of photodoping technique. b) Optical microscope image of the trilayer (device III) MoS$_2$ FET. Scale bar of 10 $\mu$m. c) Raman spectra for the several devices explored in this work. The $E^1_{2g}$ and $A_{1g}$ labels identify the two vibrational modes used to determine MoS$_2$ thickness. d) Sketch showing the time dependence of drain-source current, $I_{ds}$, with and without laser illumination, indicating reduced normalized current noise power under laser illumination.}
     \label{sketch}
   \end{figure}

\section{Experimental details}

We studied four MoS$_2$ FET devices with monolayer, bilayer, trilayer and bulk-like MoS$_2$ channels, henceforth identified as device I, II, III and IV, respectively. The sample preparation was carried out using high-purity bulk MoS$_2$ crystals acquired from a commercially available supplier (SPI Supplies). The MoS$_2$ FETs were fabricated on n$^+$-doped silicon dies with thermally grown 250-nm-thick SiO$_2$. The substrates were cleaned and sonicated in acetone, isopropanol, and deionized water and dried on a hotplate at 195 $^{\circ}$C. Following an all-dry viscoelastic stamping deterministic transfer procedure \cite{Castellanos2014} using polydimethylsiloxane (PDMS), MoS$_2$ flakes with different number of atomic layers were optically identified and selectively transferred onto the SiO$_2$ substrates. Standard electron-beam lithography using a double layer of poly methyl-methacrylate (PMMA, 495/950 MDa) was employed to pattern Ti(5 nm)/Au(35 nm) metal contacts. The metallization was done in an ultrahigh vacuum deposition system at a base pressure of $10^{-9}$ mbar, using electron-beam evaporation to deposit Ti and Au. The lift-off was performed in acetone. Figure \ref{sketch}a) shows the sketch of the FET electrical configuration, with MoS$_2$ channel, SiO$_2$(250 nm) dielectric, Si backgate, and Ti(5 nm)/Au(35 nm) contacts. Figure \ref{sketch}b) shows a microscopic picture of device III.

The thickness of the MoS$_2$ flakes was determined using Raman spectroscopy with a 532 nm laser line at room temperature (see Figure \ref{sketch}c). The difference between the frequency of the Raman $E^1_{2g}$ and $A_{1g}$ peaks, $\Delta (A_{1g}-E^1_{2g})$, can be used as a reliable indicator of the number of layers of the flake \cite{Li2012}. The frequencies of the peaks were determined using a double Lorentzian least square best fit. The obtained $ \Delta (A_{1g}-E^1_{2g}$) of 20.2, 22.8, 23.2, 24.9 cm$^{-1}$ are in agreement with the expected values for monolayers, bilayers, trilayers and bulk-like, respectively. The  lateral dimensions (length $L$ and width $W$) of device I were $L=3.7$ $\mu$m and $W=2.8$ $\mu$m; device II, $L=3.6$ $\mu$m and $W=4.3$ $\mu$m; device III, $L=5.6$ $\mu$m and $W=5.5$  $\mu$m; device IV, $L=52$ $\mu$m and $W=12$  $\mu$m.

Due to the relatively high resistance of the MoS$_2$ FETs, both current and current noise were measured with a Keithley 6485 picoammeter. A home-made battery designed for ultra-low noise at a bias of 1 V powered the voltage applied between the drain and the source contacts, while a gate voltage up to 80 V was applied between the substrate and the drain using a Keithley 228A voltage source. In the set of measurements performed during this study, the amplifier noise and Johnson noise were orders of magnitude below the detected noise-levels, which allows us to disregard spurious origins for the observed signals. 

The current intensity time series were measured at every fixed gate during 180 s with resolution of 67 ms. The FETs were illuminated with a TOPTICA-iBeam Smart diode laser with wavelength of 487 nm and with up to 1mW of nominal output power. The maximum effective light surface density was estimated to be substantially below 1 $\mu$W/$\mu$m$^2$. The laser spot of about tens of $\mu$m in diameter covered the whole surface of the MoS$_2$ channel. All the measurements were performed at room temperature and in ambient conditions. LFN experiments were carried out by studying the current relaxation time series after steeply sweeping up the gate voltage by 2 V under two different conditions: (i) dark conditions, where no laser illumination was applied, and (ii) in the presence of laser illumination. Figure \ref{sketch}d) shows schematically how the normalized current noise varies under photodoping.

\section{Results and discussion}
\subsection{Electron transport with and without photodoping}

In the body of the manuscript, we mainly concentrate on the results obtained for devices II and III, between which a significant change in the LFN under photodoping has been observed. Some relevant complementary results for devices I and IV are provided in the supplemental materials \cite{Supplemental materials}.

\begin{figure}[!h]
    \includegraphics[width=\linewidth]{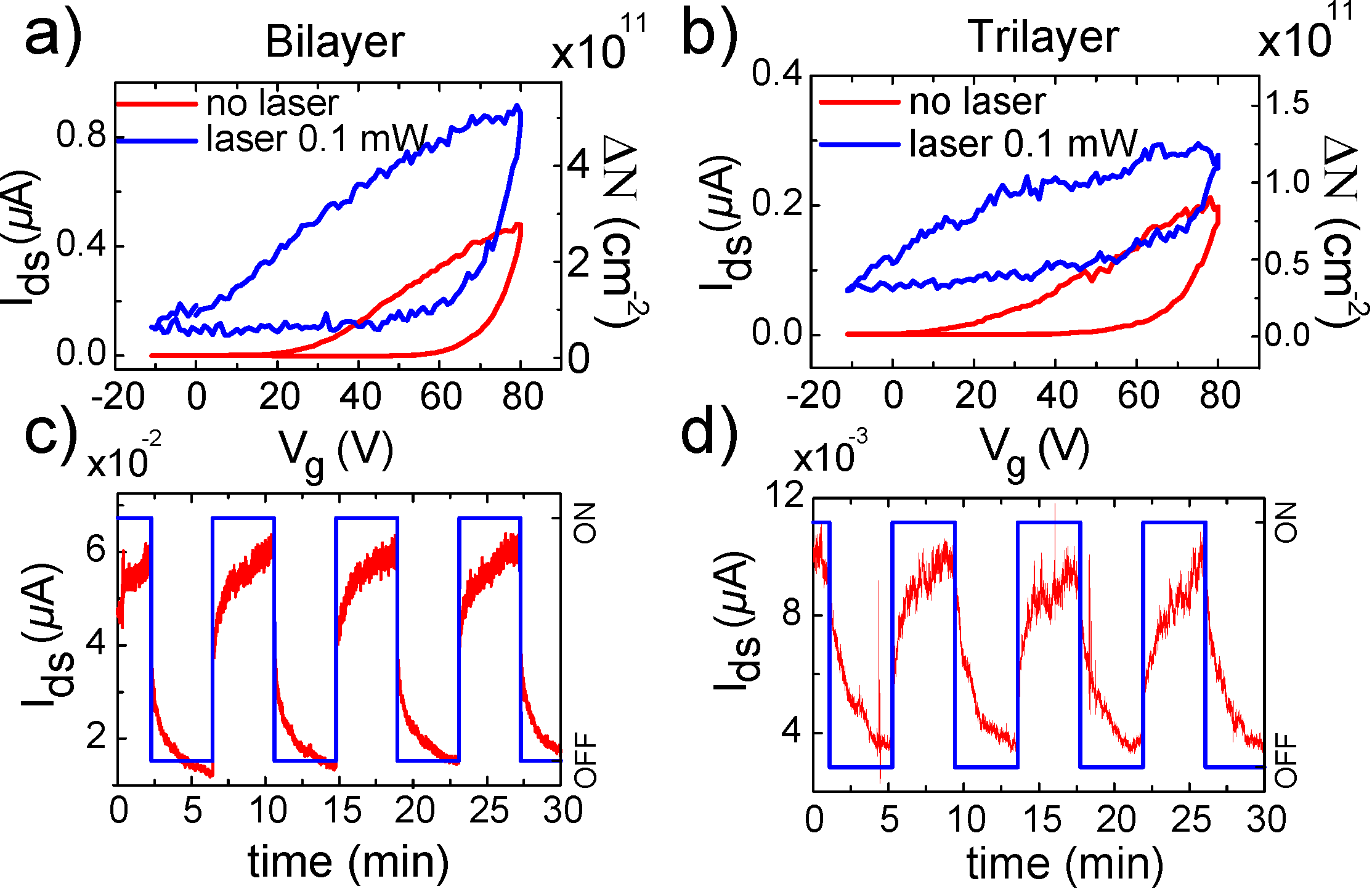}
    \centering
     \caption{Transfer characteristics of the drain-source current $\textit{I}_{ds}$ as a function of the gate voltage $\textit{V}_{g}$ with and without laser illumination of a) device II and b) device III. Gate voltage sweep rate is 1 V/s. The right axes presents the corresponding gate dependence of the photoexcited carrier concentration (see text for the details).
   Plots c) and d) show the laser pulse response of the photocurrent for devices II and III, respectively,  without applied $\textit{V}_{g}$, and applied drain-source bias $\textit{V}_{ds}$ of 1 V. Blue solid line represents the time dependence of the ON/OFF incidence of illumination with laser output power of 0.1 mW. Period T = 500 s.}
     \label{transporte}
   \end{figure}

We start by evaluating the DC transport with and without illumination to establish a foot-ground for the LFN studies. Figures \ref{transporte}a) and \ref{transporte}b) show the gate dependence of the drain-source current $\textit{I}_{ds}$ (transfer characteristics) with and without laser illumination.  The non-illuminated FETs exhibit n-type behavior with ON/OFF ratios of the order of $10^5-10^6$. In agreement with previous reports \cite{Late2012,Fontana2013}, the transfer characteristics of devices II and III reveals hysteretic behavior due to  current relaxation. Qualitatively similar effects are shown \cite{Supplemental materials} in Supplemental Figure 1S for devices I and IV. With laser illumination the current output of the transistor OFF state greatly increases from $10^{-12}$ to $10^{-7}$A,  while the current output at the ON state increases by a factor of two. In both cases the hysteretic behavior slightly increases, in line with previous studies\cite{Late2012}.

Devices II and III exhibit maximum field-effect mobilities of 9 and 14 cm$^2$/Vs respectively. The field-effect mobility of devices II and III are calculated from the expression $\mu_{FE}$ = $Lg_m/(WC_{ox}V_{ds})$, where $g_m$ is the terminal transconductance ($dI_{ds}/dV_g$), and $C_{ox}$ the gate capacitance per unit area, estimated to be 1.38$\times$10$^{-4}$F/m$^2$ for a 250-nm-thick SiO$_2$ dielectric, using a parallel plate model.  Ti/Au contacts to the MoS$_2$ channel provide low Schottky barriers of 0.05 eV \cite{Allain2015}, which at room temperature result in  ohmic-like output characteristics \cite{Supplemental materials} (see supplemental Figure 2S).

Figures \ref{transporte}c) and \ref{transporte}d) show the drain-source photocurrent generated by a laser pulse of period T = 500 s at zero gate voltage for devices II and III, respectively, under an applied drain-source bias, $\textit{V}_{ds}$ of 1 V. Exponential decay time constants of about 180 s are observed before reaching the equilibrium state, both for the ON and OFF states of the laser illumination. Similar curves for devices I and IV are shown \cite{Supplemental materials} in Supplemental Figures 3Sa) and 3Sb). The normalized variation of the pulsed photocurrent as a function of the MoS$_2$ channel thickness \cite{Supplemental materials} (Supplemental Figures 3Sc) points out at a difference between devices with increasing number of layers, with maximum photocurrent response at 2 or 3 layers. 

We estimate the additional photoexcited carrier density, $\Delta N$, generated under illumination using a simplified relation between conductance, mobility, and carrier concentration, given by  $\Delta\sigma\approx e(\mu_h \Delta N_h+\mu_e  \Delta N_e)$, with $\Delta\sigma $ being the change in surface conductance, $\mu_h $ and $\mu_e$ hole and electron mobilities, and $\Delta N_h, \Delta N_e$ the change in hole and electron concentrations. Being n-doped, we assume the presence of one type of carriers only (electrons) and that for the minimum applied laser power (0.1 mW) the carrier concentration change is larger than the change in the Hall mobility\cite{Britton2013}. Figure \ref{transporte}a) and b) show the change in carrier density between dark and illuminated states for device II and III.


\subsection{Low frequency noise without photodoping}

We now turn to investigate how the LFN parameters change along the current relaxation. For this, we used the determined exponential decay time constant of about 180 s. Photocurrent relaxation of similar time scales has been reported for MoS$_2$ FETs \cite{Wu2015}, being attributed to the presence of random potentials in the device due to defects. For the noise analysis, we divided the current time series (both in dark condition and under illumination) in three periods, $P1, P2, P3$, each lasting for about 60 $s$. For each of those intervals, the noise power spectrum of the current has been analyzed separately by using the Hooge relation \cite{Sangwan2013,Hooge1969},  $S_I= \frac{\alpha I_{ds}^2}{f^{\beta}}$, where $S_I$ is the square of the module of the fast Fourier transform of the current time series expressed as a function of the frequency $f$, the source-drain current $I_{ds}$ under equilibrium, and $\alpha$ and $\beta$ characteristic Hooge parameters obtained from fits in the $0.05-5 Hz$ frequency range. When $\beta$ is close to one (i.e. roughly between 0.7 and 1.5) the noise is usually called $1/f$ noise \cite{Hooge1969}. The strong dependence of the channel resistivity with gate voltage (with a change up to 6 orders of magnitude between the ON and OFF states), requires that we fix the applied V$_{ds}$ and record continuously $I_{ds}$ flowing through the device for consecutively increasing gate voltage steps of 2 V, with current fluctuations recorded for 180 s. Supplemental Figure 2S shows \cite{Supplemental materials} that the devices are close to the Ohmic regime for the full range of gate voltages explored in this work, where $S_I$ is proportional to $I_{ds}^2$.

\begin{figure}[!h]
   \includegraphics[width=\linewidth]{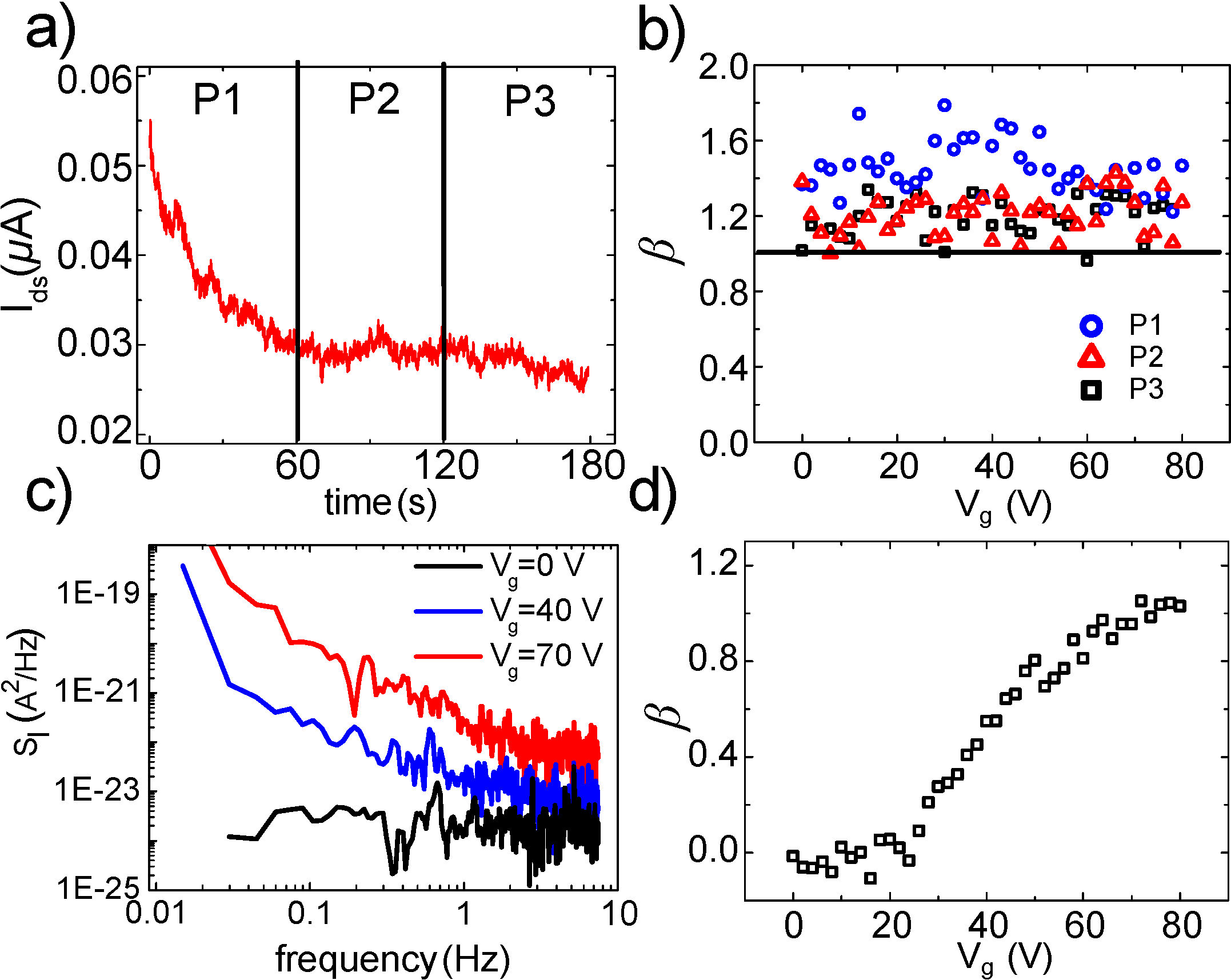}
   \centering
     \caption{a) Drain-source current $\textit{I}_{ds}$ dependence with time for device III after the gate voltage $V_g$ is steeply swept up by 2V under dark conditions. The current time series corresponds to a time window of 180 s (determined from the photocurrent exponential decay time constant), divided into three periods, P1, P2, P3, of 60 s each. b) Dependence of the characteristic Hooge parameter $\beta$ with gate voltage $V_g$ at P1, P2, and P3 for device III under dark conditions. Solid line represents the ideal $\beta$ value for $1/f$-type LFN. c) Noise power spectrum measured for device II at P3 for different gate voltages under dark conditions. d) Corresponding gate voltage dependence of $\beta$ for device II under dark conditions at P3.}
     \label{series}
   \end{figure}

Figure \ref{series}a) exhibits the current time series dependence of device III under dark conditions after a steep sweep of $V_g$ by 2 V, and \ref{series}b) the corresponding gate dependence for the extracted parameter $\beta$ at P1, P2, P3. At the P3 interval, the relaxation effects are found to be negligible, and indicate that the system is close to equilibrium. Evaluating $\beta$ as the device operation approaches equilibrium (from P1 to P3), the exponent $\beta$ decreases from above 1.5 to closer to 1 (see Figure \ref{series}b), without exhibiting a significant gate dependence. The larger $\beta$ values in the presence of time dependent relaxation were also observed in off-equilibrium magnetic tunnel junctions near the switching to the antiparallel state \cite{Guerrero2002}.  This dependence of the LFN with the period of the current time series was observed in all samples. In order to address possible concerns about spectral leakage along relaxation, we analyzed the LFN data only for the periods closest to equilibrium (P3).

Focusing henceforth on the current time series period P3, device II exhibits a clear dependence of the $S_I (f)$ with gate voltage. Figure \ref{series}c) shows the noise power spectrum under dark conditions at $V_g$= 0, 40, and 70 V for device II. Plotting $S_I (f)$ in log-scale allows us to clearly identify a power dependence with frequency with Hooge parameter $\beta$ increasing for higher gate voltages. Figure \ref{series}d) summarizes the gate dependence of $\beta$ for device II.  For $V_{g} <$ 20 V, $\beta$ is close to 0, changing to  $1/f$ noise when $V_{g} >$ 40 V. These results suggest a dependence of the LFN under dark conditions with channel thickness. Still, at sufficiently high gate and close to equilibrium (P3 time window), $\beta$ is close to 1. However, for device I and II at the OFF state $\beta$ approaches 0, suggesting that a strong random telegraph noise (RTN) overcomes $1/f$- type noise. At higher gates and with increasing number of layers, the $1/f$ noise contribution overcomes this frequency independent response at low frequencies (see Figure \ref{series} b) and d)). In view of the strong decrease of the defect-assisted recombination times with decreasing number of layers in MoS$_2$ \cite{Wang2015V1,Wang2015V2}, the absence of the $1/f$ contribution at low gates for device I and II suggests that dominating generation-recombination processes are leading to a Lorentzian spectrum, where corresponding noise power spectrum tends to constant noise power values at low frequencies. The observed behavior of $\beta$ with layer number and electrostatic gating is then likely related to the interplay between two mechanisms in the MoS$_2$ monolayer to few-layer flakes: the strongly dependent recombination time scales with number of layers \cite{Wang2015V1,Wang2015V2}, and the different intrinsic doping levels of each flake, known to vary strongly with ambient conditions \cite{Kwon2014v2}.

\subsection{Tunning different noise regimes by photodoping}

Two models based on the Hooge relation can be used to describe the origins of the fluctuations in FETs exhibiting $1/f$-type power spectral density. These are the carrier mobility fluctuations (HMF) and charge number fluctuations (CNF) models \cite{Simoen1999}. In the CNF model, $S_I  \propto (V_g-V_{th})^{-2}$, with $V_{th}$ being the threshold gate voltage for the opening of conductance channels. In the HMF model, $S_I  \propto N^{-1}$, with $N$ being the carrier density. The low conductivity of MoS$_2$ FETs places it in the limiting case of weakly conducting regimes, and therefore the CNF model should suit the LFN, where the drain-source current noise power spectral density is expected to show a quadratic dependence on the gate voltage. The HMF model is usually valid for conducting regimes \cite{Hooge1969}.

In the following, we demonstrate that the combined application of gate voltage and laser illumination to our MoS$_2$ FETs, which affects the carrier density $N$ (Figure \ref{transporte}), changes also the transport conditions from $1/f$-type noise dominated by CNF to noise dominated by HMF.

\begin{figure}[!h]
   \includegraphics[width=\linewidth]{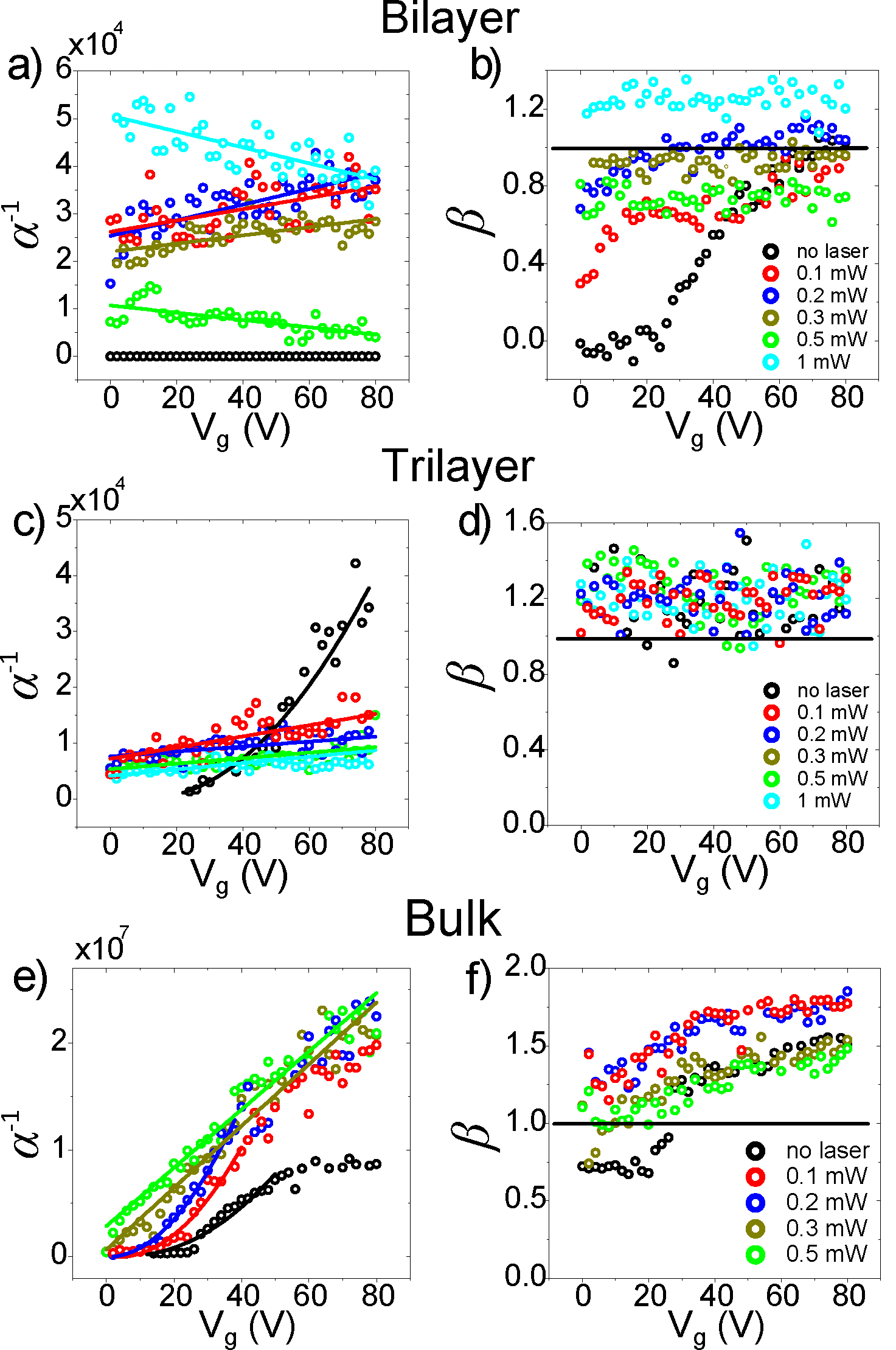}
   \centering
     \caption{Gate dependence of $\alpha^{-1}$ and $\beta$  Hooge parameters for devices II, III and IV, determined from the LFN measurements under dark conditions and with illumination. Laser power ranges from absence of illumination (dark condition referred to as "no laser") to 1mW. a) ,b) Device II. (c,d) Device III. e),f) Device IV. Solid lines in $\alpha^{-1}$ plots are fittings to the gate voltage dependence, and in $\beta$ plots represent the ideal $\beta$ value for 1/f-type LFN.}
     \label{noise}
   \end{figure}

Following Ref.\cite{Sangwan2013}, we consider overdrive conditions for the analysis, when $|V_g - V_{th}|>0$, and the carrier density $N$ can be approximated as $N \propto (V_{g} -V_{th})$.  We shall restrict our study to the cases where the power spectrum approximately follows $1/f$ behavior (0.7 $< \beta <$ 1.5). In order to reach specific conclusions on the physical processes responsible for the LFN, we plot the gate dependence of the inverse of the normalized noise parameter $\alpha$ for different photodoping conditions (Figures \ref{noise}a), c), e)). 

Starting with dark conditions, the noise measurements in device I \cite{Supplemental materials} (Supplemental Figure 4S) and device II  (Figure \ref{noise}b) suggest that the transition to a sufficiently conducting regime, with $1/f$-type noise, occurs only at gates above 30-40 V. This restricts the inverse noise ($\alpha^{-1}$ vs. $V_g$) analysis to the relatively narrow range of 40 to 80 V, where the corresponding $\alpha^{-1}$ dependence shows a linear trend, indicating HMF as the driving mechanism for LFN. However, the dispersion of the normalized noise parameters has been found to be extremely large for device I \cite{Supplemental materials} (see Supplemental Figure 4S). Devices III and IV, on the other hand,  exhibit a quadratic dependence of $\alpha^{-1}$ as a function of the gate voltage (see Figs. \ref{noise}c) and \ref{noise}e), respectively). This fact, together with the fact that in the explored gate voltage range the devices show $1/f$-type noise (from the $\beta$ parameter analysis), allows us to point at CNF as the underlying physical mechanism driving the LFN\cite{Simoen1999}. These results allow us to separate the dark condition LFN in devices I to IV into two categories. More specifically, device I and II showed $1/f$-type noise driven by HMF only above some threshold gate voltage (of about 40 V for device II and 60 V for device I \cite{Supplemental materials} (see Supplemental Figures 4S and 5S)), with mainly RTN for lower gate voltages; while devices III and IV revealed approximately $1/f$ noise driven by carrier number fluctuations in the whole range of applied gates from 0 to 80 V (see Figure \ref{noise}).

Under illumination, devices I and II show a linear dependence of $\alpha^{-1} \propto (V_{g}-V_{th})$. With increased photodoping, the dependence of $\beta$ with the gate voltage changes from RTN-type, (seen under dark conditions for gate voltages lower than 40 V), to $1/f$-type across the full gate range, demonstrating how the additional photoconducting channels contribute to LFN dominated by HMF \cite{Hooge1969}. Interestingly, in device II, for the maximum illumination power of 1mW,  the drain-source current dependence with gate voltage exhibits a decrease for higher gate voltages \cite{Supplemental materials} (Supplemental Figure 6S shows this particular case), as well as a strong reduction of the normalized noise power (Fig.\ref{noise}a)). We tentatively attribute this effect to an increase in the recombination rates of the charge carriers leading to a decrease in conductance and increase in LFN. For devices III and IV, one clearly observes a transition from a quadratic dependence $\alpha^{-1} \propto (V_{g}-V_{th})^2$ to a linear dependence  $\alpha^{-1} \propto (V_{g}-V_{th})$, pointing at a crossover from carrier number fluctuations\cite{Simoen1999} to fluctuations in the carrier mobility \cite{Hooge1969}, respectively. Note that the LFN data obtained for $V_g$ below $V_{th}$ (\textit{i.e.} practically in the OFF state) is not used in our fits to extract $V_{th}$, since the McWhorter model is valid only for overdrive conditions, where $V_g >V_{th}$. Table \ref{Table} summarizes the several observations under dark and illumination conditions, electrostatic gating (for overdrive conditions), and different layer count.

\begin{table}[ht]
\centering
\caption{Summary of the LFN for devices I, II, III and IV, under dark conditions ("Dark") and under laser illumination ("Light"), for two extreme electrostactic gating regimes (low gate voltage, but still above overdrive conditions, V$_g$ $\geq$ V$_{Th}$, and at the highest gate voltages, V$_g$ $\gg$ V$_{Th}$.\newline}
\begin{tabular}{|c|c|c|c|c|}
\hline
 & \multicolumn{2}{|c|}{I, II}  & \multicolumn{2}{|c|}{III, IV}  \\
\hline
& V$_g$ $\geq$ V$_{th}$  & V$_g$ $\gg$ V$_{th}$ & V$_g$ $\geq$ V$_{th}$  & V$_g$ $\gg$ V$_{th}$ \\
\hline
Dark & RTN  & HMF & CNF  & CNF \\
\hline
Light & HMF  & HMF & HMF  & HMF \\
\hline
\end{tabular}
\label{Table}
\end{table}

The observation of RTN for devices I and II under dark conditions (and low gate voltage), while devices III and IV at the same conditions show CNF, was already discussed in section B, with the effect originating from the strong dependence of the recombination time-scales for different layer numbers \cite{Wang2015V1,Wang2015V2}. At low gate voltages, by exciting device I and II with light, the carrier density enhancement is enough to drive 1/$f$ noise, also of HMF character.

The observed crossover from CNF to HMF in device III and IV under laser illumination can be understood as a consequence of the percolative character of conductance in the random resistor network of MoS$_2$ FETs \cite{Paul2016}. Within this picture, the change in the noise microscopic mechanism driven by light from CNF to HMF reflects the crossing from a regime where transport happens via hopping or tunneling between disconnected metallic puddles at the fermi-level, in the so called "island-and-sea" representation of the carrier distribution in the MoS$_2$ flakes, to a continuum percolation, where instead of a random network of disconnected metallic puddles there is a continuum electron sea at the fermi-level.

Further analysis of the table allows us to extract one additional observation. Under sufficient electron-doping (photodoping and high electrostatic gating), all devices show HMF. Together with the fact that electrostatic gating under dark conditions is not enough to drive devices III and IV to HMF LFN (while device I and II show HMF), strongly suggests that the intrinsic doping levels of devices I and II are higher than in devices III and IV. This observation is compatible with other reports where they observe a higher carrier density for thinner devices\cite{Britton2013}.
   \begin{figure}[!h]
   \includegraphics[width=0.8\linewidth]{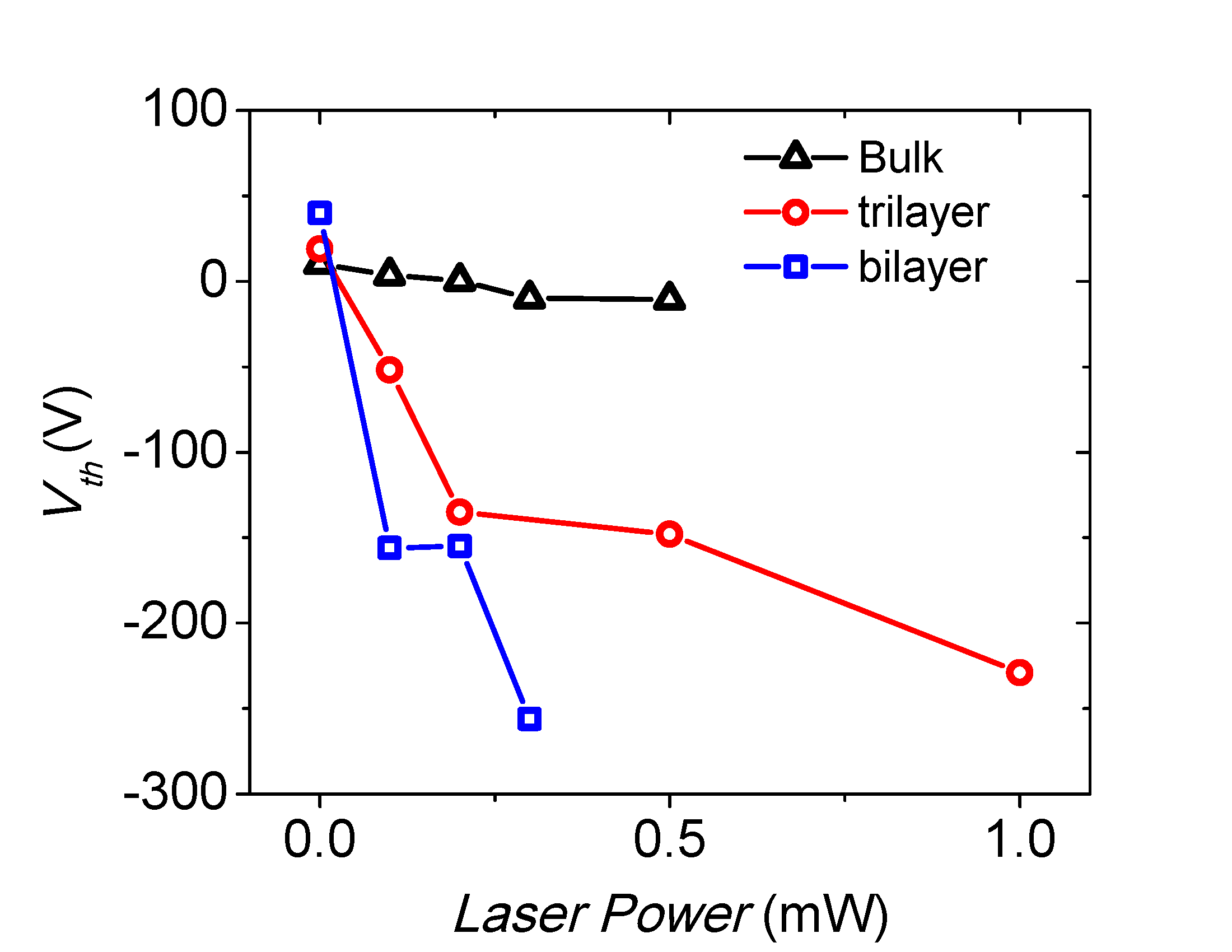}
     \centering
     \caption{Estimated threshold voltage for devices II, III and IV as a function of the laser illumination power. The threshold voltage is determined for a 250-nm-thick SiO$_2$ dielectric.}
     \label{Threshold}
   \end{figure}

A comparison of the influence of the illumination power on $V_{th}$ (Figure \ref{Threshold})  shows that thinner devices require lower illumination power to achieve significant photodoping, which is reflected by the shift of $V_{th}$ to lower values. $V_{th}$ is determined as the equivalent field-effect electrostatic doping required, with the use of a 250-nm-thick SiO$_2$ dielectric, to change the MoS$_2$ device operation from an insulating to a more conductive regime when under the specified illumination conditions (photodoping).

Gathering the several conclusive elements for the dependence of $\beta$ and $\alpha^{-1}$ with the layer thickness, gating, and laser illumination, the observed effects arise then from a complex interplay between the several electron-doping sources (the intrinsic doping of the flakes, the photodoping, and the electrostatic doping), and the strong decrease of the recombination time scales with decreasing layer number. RTN noise gives place to 1/$f$-type HMF noise in the thinner devices (I and II) under sufficient electron-doping, while the thicker devices (III and IV) exhibit mainly 1/$f$-type CNF for overdrive conditions. When in the 1/$f$-type noise regime, we observe a crossover from CNF to HMF driven by photodoping, where by shining light the discontinuous random resistive network percolates to a continuum electron sea at the fermi level.

\section{Conclusions}

We have introduced a method to access different LFN regimes in MoS$_2$ FETs by using photodoping, and identified a constraint in the transient decay time window to perform LFN studies.  We believe this concept to be applicable in a much wider class of 2D materials-based FETs. Our results confirm the presence of frequency independent low-frequency generation-recombination noise, previously reported for monolayer MoS$_2$ FETs studied under environmental conditions\cite{Sangwan2013}, attributed to either traps in the SiO$_2$ substrate or to midgap defect states in MoS$_2$For sufficiently thick MoS$_2$ FETs (above bilayer), electrostatic gating in the dark state reveals $1/f$-type noise driven by carrier number fluctuations, in agreement with most of the previous reports \cite{Joo2016,Ko2016,Das2015,Lin2015,Sharma2014,Ghatak2014,Xie2014,Renteria2014,Kwon2014,Na2014,Sharma2015,Rumyantsev2015}.
In these conditions, by using photodoping with reasonably small laser powers, we are able to tune the origin of the conductance fluctuations from carrier number fluctuations to carrier mobility fluctuations. Our findings introduce then a versatile approach to investigate LFN in 2D vdW-based FETs, paving the way to overcome LFN limitations of TMD-based photodetectors and transistors.

\section*{ACKNOWLEDGEMENTS}
The work in Madrid has been supported by Spanish MINECO (MAT2015-66000-P)  and the Comunidad de Madrid through NANOFRONTMAG-CM (S2013/MIT-2850) grants. The work in CIC nanoGUNE has been supported by the European Union 7th Framework Programme under the Marie Curie Actions (607904-13-SPINOGRAPH), by the Spanish MINECO under Project No. MAT2015-65159-R, and by the Basque Government under Project No. PC2015-1-01.


\begin{thebibliography} {23}

\bibitem{Ambrosetti2016}
A. Ambrosetti, N. Ferri, R. A. DiStasio Jr., A. Tkatchenko, Wavelike charge density fluctuations and van der Waals interactions at the nanoscale,  Science \textbf{351}, 1166 (2016).

\bibitem{Balandin2013} 
A. Balandin, Low-frequency 1/f noise in graphene devices, Nat. Nanotechnol. \textbf{8}, 549 (2013).


\bibitem{Sangwan2013}
V. K. Sangwan, N. H. Arnold, D. Jariwala, T. J. Marks, L. J. Lauhon, M. C. Hersam, Low-Frequency Electronic Noise in Single-Layer MoS$_2$ Transistors,  Nano Lett. \textbf{13}, 4351 (2013).

\bibitem{Renteria2014}
J. Renteria, R. Samnakay, S. L. Rumyantsev, C. Jiang, P. Goli, M. S. Shur, A. A. Balandin, Low-frequency 1/f noise in MoS$_2$ transistors: Relative contributions of the channel and contacts, Appl. Phys. Lett. \textbf{104}, 153104 (2014).

\bibitem{Kwon2014}
H-J. Kwon, H. Kang, J. Jang, S. Kim, C. P. Grigoropoulos, Analysis of flicker noise in two-dimensional multilayer MoS$_2$ transistors, Appl. Phys. Lett. \textbf{104}, 083110 (2014).

\bibitem{Cho2015} 
I.-T. Cho, J. I. Kim, Y. Hong, J. Roh, H. Shin, G. W. Woo Baek, C. Lee, B. H. Hong, S. H. Jin, J-H. Lee,  Low frequency noise characteristics in multilayer WSe$_2$ field effect transistor, Appl. Phys. Lett. \textbf{106}, 023504 (2015).

\bibitem{Radisavljevic2015}
B. Radisavljevic, A. Radenovic, J. Brivio, V. Giacometti, A. Kis, Single-layer MoS$_2$ transistors , Nat. Nanotechnol. \textbf{6}, 147 (2011).

\bibitem{Sarkar2015}
D. Sarkar, X. Xie, W. Liu, W. Cao, J. Kang, Y. Gong, S. Kraemer, P.M. Ajayan, K. Banerjee,  A subthermionic tunnel field-effect transistor with an atomically thin channel, Nature \textbf{526}, 91 (2015).

\bibitem{Wang2012} 
H. Wang, L. Yu, Y.-H. Lee, Y. Shi, A. Hsu, M. L. Chin, L. J. Li, M. Dubey, J. Kong, T. Palacios, Integrated Circuits Based on Bilayer MoS$_2$ Transistors, Nano Lett. \textbf{12}, 4674 (2012).

 \bibitem{Kim2012}
 S. Kim, A. Konar, W.-S. Hwang, J. H. Lee, J. Lee, J. Yang, C. Jung, H. Kim, J.-B. Yoo, J.-Y Choi,  High-mobility and low-power thin-film transistors based on multilayer MoS$_2$ crystals, Nat. Commun. \textbf{3}, 1011 (2012).

\bibitem{Furchi2014}
M. M. Furchi, A. Pospischil, F. Libisch, J. Burgd\"{o}rfer, T. Mueller, Photovoltaic Effect in an Electrically Tunable van der Waals Heterojunction,  Nano Lett. \textbf{14}, 4785 (2014).

\bibitem{Memaran2015} 
S. Memaran, N. R. Pradhan, Z. Lu, D. Rhodes, J. Ludwig, Q. Zhou, O. Ogunsolu, P. M. Ajayan, D. Smirnov, A. I. Fern\'andez-Dominguez, F. J. Garc\'ia-Vidal, L. Balicas, Pronounced Photovoltaic Response from Multilayered Transition-Metal Dichalcogenides PN-Junctions, Nano Lett. \textbf{15}, 7532 (2015).

\bibitem{Buscema2015} 
M. Buscema, J. O. Island, D. J. Groenendijk, S. I. Blanter, G. A. Steele, H.S. J. van der Zant, A. Castellanos-Gomez, Photocurrent generation with two-dimensional van der Waals semiconductors, Chem. Soc. Rev. \textbf{44}, 3691 (2015).

\bibitem{Perkins2013}
F. K. Perkins, A. L. Firedman, E. Cobas, P. M. Campbell, G. G. Jernigan, B. T. Jonker, Chemical Vapor Sensing with Mono layer MoS$_2$, Nano Lett. \textbf{13}, 668 (2013).

\bibitem{Lin2015} Y-F. Lin, Y. Xu, C-Y Lin, Y-W. Suen, M. Yamamoto, S. Makaharai, K. Ueno, K. Tsukagoshi, Origin of Noise in Layered MoTe2 Transistors and its Possible Use for Environmental Sensors, Adv. Mater. \textbf{27}, 6612 (2015).

\bibitem{Late2012}
Late, D. J., Liu, B., Matte, H. S. S. R., Dravid, V. P., Rao, C. N. R.,  Hysteresis in Single-Layer MoS$_2$ Field Effect Transistors,  ACS Nano \textbf{6}, 5635 (2012).

\bibitem{Guo2015}
Y. Guo, X. Wei, J. Shu, B. Liu, J. Yin, C. Guan, Y. Han, S. Gao, Q. Chen, Charge trapping at the MoS$_2$-SiO$_2$ interface and its effects on the characteristics of MoS$_2$ metal-oxide-semiconductor field effect transistors, Appl. Phys. Lett. \textbf{106}, 103109 (2015).

\bibitem{Wu2015} 
Y-C. Wu, C-H. Liu, S-Y. Chen, F-Y. Shih, P-H. Ho, C-W. Chen, C-T. Liang, W-H. Wang, Extrinsic Origin of Persistent Photoconductivity in Monolayer MoS$_2$ Field Effect Transistors, Sci. Rep. \textbf{5}, 11472 (2015).

\bibitem{Joo2016}
M.-K. Joo, Y. Yun, S. Yun, Y. H. Lee, D. Suh, Strong Coulomb scattering effects on low frequency noise in monolayer WS$_2$ field-effect transistors, Appl. Phys. Lett. \textbf{109}, 153102 (2016).

\bibitem{Ko2016}
S.-P. Ko, J.M. Shin, Y.J. Kim, H.K. Jang, J.E. Jin, M. Shin, Y.K. Kim, G.T. Kim, Current fluctuation of electron and hole carriers in multilayer WSe$_2$ field effect transistors, Appl. Phys. Lett. \textbf{107}, 242102 (2015).

\bibitem{Das2015}
S. R. Das, J. Kwon, A. Prakash, C. J. Delker, S. Das, D. B. Janes, Low-frequency noise in MoSe$_2$ field effect transistors, Appl. Phys. Lett. \textbf{106}, 83507 (2015).

\bibitem{Sharma2014} D. Sharma, M. Amani, A. Motayed, P. B. Shah, A. G. Birdwell, S. Najmaei, P. M. Ajayan, J. Lou, M. Dubey, Q. Li,  A. V. Davydov, Electrical transport and low-frequency noise in chemical vapor deposited single-layer MoS$_2$ devices, Nanotechnology \textbf{25}, 155702 (2014).

\bibitem{Ghatak2014}
S. Ghatak, S. Mukherjee, M. Jain, D. D. Sarma, A. Ghosh,  Microscopic origin of low frequency noise in MoS$_2$ field-effect transistors,  APL Materials \textbf{2}, 092515 (2014).

\bibitem{Xie2014}
X. Xie, D. Sarkar, W. Liu, J. Kang, O. Marinov, M. J. Deen, K. Banerjee, Low-Frequency Noise in Bilayer MoS$_2$ Transistor,  ACS Nano \textbf{8}, 5633 (2014).

\bibitem{Na2014} 
J. Na, M.-K. Joo, M. Shin, J. Huh, J.-S. Kim, M. Piao, J.-E. Jin, H.-K. Jang, H. J. Cho, J. H. Shimd, G.-T. Kim, Low-frequency noise in multilayer MoS$_2$ field-effect transistors: the effect of high-k passivation Nanoscale \textbf{6}, 433 (2014).

\bibitem{Sharma2015}
D. Sharma, A. Motayed, P. B. Shah, M. Amani, M. Georgieva, A. G. Birdwell, M. Dubey, Q. Li, A. V. Davydov, Transfer characteristics and low-frequency noise in single-and multi-layer MoS$_2$ field-effect transistors, Appl. Phys. Lett. \textbf{107}, 162102 (2015).

\bibitem{Rumyantsev2015}
S. L. Rumyantsev, C. Jiang, R. Samnakay, M. S. Shur, A. A. Balandin, 1/ f Noise Characteristics of MoS$_2$ Thin-Film Transistors: Comparison of Single and Multilayer Structures,  IEEE Electron Device Lett. \textbf{36}, 517 (2015).

\bibitem{Hooge1969}
F. N. Hooge, 1/F noise is no surface effect, Phys. Lett. \textbf{A29}, 139 (1969).

\bibitem{Paul2016} 
T. Paul, S. Ghatak,  A. Ghosh, Percolative switching in transition metal dichalcogenide field-effect transistors at room temperature, Nanotechnology \textbf{27}, 125706 (2016).

\bibitem{Castellanos2014} 
A. Castellanos-Gomez, M. Buscema, R. Molenaar, V. Singh, L. Janssen, H. S. J. van der Zant, G. A. Steele, Deterministic transfer of two-dimensional materials by all-dry viscoelastic stamping, 2D Mater. \textbf{1}, 011002 (2014).

\bibitem{Li2012} 
H. Li, Q. Zhang, C. C. R. Yap, B. K. Tay, T. H. T. Edwin, A. Oliver, D. Baillargeat, From Bulk to Monolayer MoS$_2$: Evolution of Raman Scattering, Adv. Funct. Mater. \textbf{22}, 1385 (2012).

\bibitem{Supplemental materials} see Supplemental Material at [URL will be inserted by publisher] for transfer curves of the devices I and IV; pulsed photocurrent response as a function of $MoS_2$ thickness; noise and photocurrent characterization of the single layer $MoS_2$ FET; and photocurrent response for devices I-IV at different laser powers.


\bibitem{Fontana2013}
M. Fontana, T. Deppe, A. K. Boyd, M. Rinzan, A. Y. Liu, M. Paranjape, P. Barbara, Sci. Rep. \textbf{3}, 1634 (2013).

\bibitem{Allain2015}
A.	Allain, J. Kang, K. Banerjee, A. Kis, Electrical contacts to two-dimensional semiconductors, Nat. Mater. \textbf{14}, 1195 (2015).

\bibitem{Britton2013}
Baugher, B. W. H., Churchill, H. O. H., Yang, Y. Jarillo-Herrero, P., Intrinsic Electronic Transport Properties of High-Quality Monolayer and Bilayer MoS$_2$, Nano Lett. \textbf{13}, 4212 (2013).

\bibitem{Guerrero2002} 
R. Guerrero, F. G. Aliev, R. Villar, R. Ortega-Hertogs, W. K. Park, J. S. Moodera, Low frequency noise in Co/Al$_2$O$_3$delta(Fe)/Ni80Fe20 magnetic tunnel junctions, J. Phys. D: Appl. Phys. \textbf{35} 1761 (2002).

\bibitem{Wang2015V1}
H. Wang,  C. Zhang, and F. Rana, Ultrafast Dynamics of Defect-Assisted Electron-Hole Recombination in Monolayer MoS$_2$, Nano Lett. \textbf{15}, 339–345 (2015).


\bibitem{Wang2015V2}
H. Wang, C. Zhang, and F. Rana, Surface Recombination Limited Lifetimes of Photoexcited Carriers in Few-Layer Transition Metal Dichalcogenide MoS$_2$, Nano Lett. \textbf{15}, 8204 (2015).

\bibitem{Kwon2014v2}
H.-J. Kwon, J. Jang, S. Kim,  V. Subramanian, C. P. Grigoropoulos, Electrical characteristics of multilayer MoS$_2$ transistors at real operating temperatures with different ambient conditions, Appl. Phys. Lett. \textbf{105}, 152105 (2014).

\bibitem{Simoen1999} 
E. Simoen, C. Claeys, On the flicker noise in submicron silicon MOSFETs, Solid State Electron. \textbf{43}, 865 (1999).

\end{thebibliography}
\end{document}